\documentclass[letterpaper, 10 pt, conference]{ieeeconf}
    
    \IEEEoverridecommandlockouts 
    \makeatletter

    \let\proof\@undefined
    \let\endproof\@undefined
    \makeatother
    \let\labelindent\relax 

    \usepackage{jhoagg}
    \usepackage{mathtools}
    \usepackage{amsfonts}
    \usepackage{amssymb,latexsym}
    \usepackage[psamsfonts]{eucal}
    \usepackage{amsthm}
    \usepackage{graphicx}
    \usepackage[inline]{enumitem}
    \usepackage{indentfirst}
    \usepackage{setspace}
    \usepackage{microtype}
    \usepackage{threeparttable}
    \usepackage{float}
    \usepackage{cleveref}
    \usepackage{afterpage}
    \usepackage{placeins}
    \usepackage{cancel}
    \usepackage{leftidx}
    \usepackage{breqn}
    \usepackage[export]{adjustbox}
    \usepackage{comment}
    \usepackage[ruled, linesnumbered, lined]{algorithm2e}
    \usepackage[dvipsnames]{xcolor}
    \usepackage{xcolor}
    \usepackage[most]{tcolorbox}
    \usepackage{soul}
    \usepackage[noadjust]{cite}
    \usepackage{standalone}

    \usepackage{tikz}
    \usetikzlibrary{shapes,arrows,positioning,fit}

    \usepackage{pgf}
    \usepackage{pgfplots}
    \usepgflibrary{plothandlers}
    \usepgfplotslibrary{external} 
    \pgfkeys{/pgf/number format/.cd,1000 sep={}}  
    \usetikzlibrary{calc,shapes.misc,plotmarks}
    \pgfplotsset{compat=1.13}
    
    \let\originalleft\left
    \let\originalright\right
    \renewcommand{\left}{\mathopen{}\mathclose\bgroup\originalleft}
    \renewcommand{\right}{\aftergroup\egroup\originalright}
    
    \usepackage{fancyhdr}
    \fancypagestyle{firststyle}
    {\fancyfoot{} 
    \fancyfoot[L]{\textit{Preprint submitted to 2024 American Control Conference (ACC)}}
     
    }
    
    \newcounter{thm} 
    \newtheorem{theorem}[thm]{\indent Theorem}
    
    \newtheorem{assumption}{\indent Assumption}
    
    \newtheorem{proposition}{\indent Proposition}
    
    \newtheorem{lemma}{\indent Lemma}
    
    \newtheorem{remark}{\indent Remark}
    
    \newtheorem{corollary}{\indent Corollary}
    
    \newtheorem{definition}{\indent Definition}
    
    \newtheorem{example}{\indent Example}
    
    \newtheorem{fact}{\indent Fact}
    
    \newtheorem{conjecture}{\indent Conjecture}
    
    \newtheorem{experiment}{\indent Experiment}

    \allowdisplaybreaks

    \newlist{enumA}{enumerate}{1}
    \setlist[enumA,1]{label=(A\arabic*),leftmargin=1cm}

    \newlist{enumC}{enumerate}{1}
    \setlist[enumC,1]{label=(C\arabic*),leftmargin=1cm}

    		\newcommand\xqed[1]{%
      \leavevmode\unskip\penalty9999 \hbox{}\nobreak\hfill
      \quad\hbox{#1}}
    \newcommand\exampletriangle{\xqed{$\triangle$}}

    \usepackage{cite}
    \usepackage{accents}
    
    \newlength\figureheight 
    \newlength\figurewidth

    \allowdisplaybreaks
    
    \graphicspath{ {Figures/} }

    \DeclareMathAlphabet{\mathcal}{OMS}{cmsy}{m}{n} 

    \crefname{equation}{}{}
    
    \usepackage[ruled, linesnumbered, lined]{algorithm2e}
    \SetKwInput{KwInit}{Initialize}
    \SetKwFunction{SafeSet}{SafeSet}
    \SetKwFunction{GetSafe}{GetSafe}
    \SetKwFunction{GetUnsafe}{GetUnsafe}
    
    \SetCommentSty{mycommfont}
    
    \newlist{enumalph}{enumerate}{1}
    \setlist[enumalph]{label=\textit{(\alph*)}}

\begin{document}
\title{Composition of Control Barrier Functions With Differing Relative Degrees for Safety Under Input Constraints}
\author{Pedram Rabiee and Jesse B. Hoagg
\thanks{P. Rabiee and J. B. Hoagg are with the Department of Mechanical and Aerospace Engineering, University of Kentucky, Lexington, KY, USA. (e-mail: pedram.rabiee@uky.edu, jesse.hoagg@uky.edu).}
\thanks{This work is supported in part by the National Science Foundation (1849213, 1932105) and Air Force Office of Scientific Research (FA9550-20-1-0028).}
}
                                                                                                                                                                                                                                                                                                                                                                                                                                                                                                                                                                                                                                                                                                                                                                                                                                                                                                                                                                                                                                                                                                                                                                                                                    
\maketitle

\begin{abstract}
This paper presents a new approach for guaranteed safety subject to input constraints (e.g., actuator limits) using a composition of multiple control barrier functions (CBFs).
First, we present a method for constructing a single CBF from multiple CBFs, which can have different relative degrees. 
This construction relies on a soft minimum function and yields a CBF whose $0$-superlevel set is a subset of the union of the $0$-superlevel sets of all the CBFs used in the construction.
Next, we extend the approach to systems with input constraints. 
Specifically, we introduce control dynamics that allow us to express the input constraints as CBFs in the closed-loop state (i.e., the state of the system and the controller). 
The CBFs constructed from input constraints do not have the same relative degree as the safety constraints. 
Thus, the composite soft-minimum CBF construction is used to combine the input-constraint CBFs with the safety-constraint CBFs. 
Finally, we present a feasible real-time-optimization control that guarantees that the state remains in the $0$-superlevel set of the composite soft-minimum CBF.
We demonstrate these approaches on a nonholonomic ground robot example.
\end{abstract}

\section{Introduction}\label{sec:introduction}

Control barrier functions (CBFs) are a set-theoretic approach to obtain forward invariance and guarantee that constraints are satisfied \cite{ames2016control}. 
CBFs are often integrated into real-time optimization-based control methods (e.g., quadratic programs) as safety filters \cite{rakovic2004computation,chen2018data} and in conjunction with stability constraints or performance objectives \cite{mitchell2005time}. 
CBF methods have been demonstrated in a variety of applications, including mobile robots \cite{nguyen2015safety}, unmanned aerial vehicles \cite{gillula2014sampling}, and autonomous driving \cite{ames2016control,jin2018adaptive}.
Nevertheless, verifying a candidate CBF (i.e., confirming that it satisfies the conditions to be a CBF) can be challenging \cite{wisniewski2015converse}.

For systems without input constraints (e.g., actuator limits), a candidate CBF can often be verified provided that it satisfies certain structural assumptions (e.g., constant relative degree) \cite{ames2016control}.
In contrast, verifying a candidate CBF under input constraints can be challenging, and this challenge is exacerbated in the case where the desired forward invariant set (e.g., the safe set) is described using multiple CBFs. 
Offline sum-of-squares optimization methods can be used to verify a candidate CBF \cite{wang2018,clark2021,isaly2022feasibility,pond2022fast}.  


An online approach to obtain forward invariance (e.g., safety) with input constraints is to use a prediction of the system trajectory into the future under a backup control \cite{gurrietScalableSafety2020,rabiee2023softmin,rabiee2023softmax}. 
For example,~\cite{rabiee2023softmin,rabiee2023softmax} determine a control forward invariant subset of the safe set by using a finite-horizon prediction of the system under a backup control, and to use a soft-minimum function to construct a single composite barrier function from the multiple barrier functions that arise from using a prediction horizon.
In addition, \cite{rabiee2023softmax} allows for multiple backup controls, which helps to enlarge the verified forward-invariant subset of the safe set. 
However, all of these methods rely on a prediction of the system trajectories into the future, which can be computationally intensive.


Another approach to address forward invariance subject to input constraints is to compose a single CBF from multiple CBFs. 
For example, \cite{black2023consolidated} presents a CBF composition that uses adaptable weights. 
However, the feasibility of the composition-weight update law relies on the assumption that an underlying optimization problem satisfies Slater's condition. 


This paper presents a new approach for guaranteed safety subject to input constraints (e.g., actuator limits).
The approach relies on a composite soft-minimum CBF that combines multiple safety constraints and multiple input constraints into a single CBF constraint that can be enforced through a real-time optimization, specifically, a quadratic program. 
This paper presents several new contributions. 
First, we present a method for constructing a single CBF from multiple CBFs, where each CBF in the composition can have different relative degree.
The construction relies on a soft minimum function and yields a CBF whose $0$-superlevel set is a subset of the union of the $0$-superlevel sets of all the CBFs used in the construction.

Next, we use the composite soft-minimum CBF construction to address guaranteed safety under input constraints. 
To do this, we introduce control dynamics such that the control signal is an algebraic function of the internal controller states and the input to the control dynamics is the output of a real-time optimization. 
Other methods using control dynamics and CBFs include \cite{xiao2022control,ames2020integral}.
The use of control dynamics allows us to express the input constraints as CBFs in the closed-loop state (i.e., the state of the system and the controller).
Notably, the input constraint CBFs do not have the same relative degree as the safety CBFs. 
However, this difficult is addressed using the new composite soft-minimum CBF construction. 
Then, we present a real-time-optimization control that minimizes a quadratic performance cost subject to the constraint that the state remains in the $0$-superlevel set of the composite soft-minimum CBF, which guarantees both safety and the input constraints are satisfied. 
Finally, we demonstrate the method on a simulation of a nonholonomic ground robot subject to position constraints, speed constraints, and input constraints--none of which have the same relative degree.

\section{Notation}
The interior, boundary, and closure of the set $\SA \subseteq \BBR^n$ are denoted by $\mbox{int }\SA$, $\mbox{bd }\SA$, $\mbox{cl }\SA$, respectively.
Let $\BBP^{n}$ denote the set of symmetric positive-definite matrices in $\BBR^{n \times n}$. 

Let $\eta:\BBR^n \to \BBR^\ell$, be continuously differentiable. Then, $\eta^\prime :\BBR^n \to \BBR^{\ell \times n}$ is defined as $\eta^\prime(x) \triangleq \pderiv{\eta(x)}{x}$. The Lie derivatives of $\eta$ along the vector fields of $f:\BBR^n \to \BBR^n$ and $g:\BBR^n \to \BBR^{n \times m}$ are defined as $L_f \eta(x) \triangleq \eta^\prime(x) f(x), L_g \eta(x) \triangleq \eta^\prime(x) g(x)$, and for all positive integer $d$, define $L_f^d \eta(x) \triangleq L_f L_f^{d-1} \eta(x)$.
Throughout this paper, we assume that all functions are sufficiently smooth such that all derivatives that we write exist and are continuous.


Let $\rho>0$, and consider the function $\mbox{softmin}_\rho : \BBR \times \cdots \times \BBR \to \BBR$ defined by 
\begin{equation}\label{eq:softmin}
\mbox{softmin}_\rho (z_1,\ldots,z_N) \triangleq -\frac{1}{\rho}\log\sum_{i=1}^Ne^{-\rho z_i},
\end{equation}
which is the \textit{soft minimum}. The next result relates the soft minimum to the minimum.
\begin{fact} \label{fact:softmin_bound}
\rm{
Let $z_1,\ldots, z_N \in \BBR$. 
Then,
\begin{align*}
 \min \, \{ z_1,\ldots, z_N \} - \frac{\log N }{\rho} 
 &\le \mbox{softmin}_\rho(z_1,\ldots,z_N) \\
 &< \min \, \{z_1,\ldots, z_N\}.
\end{align*}
}
\end{fact}

\section{Composite Soft-Minimum CBF} \label{sec:softmin_hocbf}

This section presents a method for constructing a single CBF from multiple CBFs, which can have different relative degrees. 
Next, this section presents a feasible real-time-optimization control that guarantees safety with respect to all of the original CBFs. 

\subsection{Problem Formulation}
Consider 
\begin{equation}\label{eq:dynamics}
\dot x(t) = f(x(t))+g(x(t)) u(t),
\end{equation}
where $x(t) \in \BBR^{n}$ is the state, $x(0) = x_0 \in \BBR^n$ is the initial condition, and $u(t) \in \BBR^m$ is the control. 

Let  $h_1,h_2,\ldots,h_\ell \colon \BBR^n \to \BBR$ be continuously differentiable, and define the \textit{safe set}
\begin{equation}\label{eq:S_s}
\SSS_\rms \triangleq \{x \in \BBR^n \colon h_{1}(x)\geq 0, h_2(x)\ge 0 \ldots, h_{\ell}(x) \geq 0 \}.
\end{equation}
Unless otherwise stated, all statements in this section that involve the subscript $j$ are for all $j \in \{ 1,2,\ldots,\ell\}$.
We assume that there exists a positive integer $d_j$ such that:
\begin{enumA}

    \item For all $x \in \SSS_\rms$ and all $i \in \{0, 1, \ldots, d_j-2\}$, $L_g L_f^i h_j(x) = 0$. 
    \label{cond:hocbf.a}
    
    \item For all $x\in \SSS_\rms, L_g L_f^{d_j-1}h_j(x) \neq 0$. 
    \label{cond:hocbf.b}
\end{enumA}
Assumptions \ref{cond:hocbf.a} and \ref{cond:hocbf.b} imply $h_j$ has well-defined relative degree $d_j$ with respect to~\eqref{eq:dynamics}; however, these relative degrees need not be equal.

Next, consider the cost function $J \colon \BBR^n \times \BBR^m\to \BBR$ defined by
\begin{equation}\label{eq:quad_cost}
    J(x, u) \triangleq \frac{1}{2}u^\rmT Q(x) u + c(x)^\rmT u,
\end{equation}
where $Q:\BBR^n \to \BBP^{m}$ and $c:\BBR^n\to \BBR^m$.
The objective is to design a full-state feedback control $u:\BBR^n \to \BBR^m$ such that for all $t \ge 0$, $J(x(t), u(t))$ is minimized subject to the safety constraint that $x(t) \in \SSS_\rms$.

\subsection{Composite Soft-Minimum CBF Construction and Control}

For $i\in\{0,\ldots,d_j-2\}$, let $\alpha_{j,i} \colon \BBR \to \BBR$ be an extended class-$\SK$ function, and consider $b_{j, i+1}:\BBR^n \to \BBR$ defined by
\begin{equation}\label{eq:b_def}
    b_{j,i+1}(x) \triangleq L_f b_{j,i}(x) + \alpha_{j,i}(b_{j,i}(x)),
\end{equation}
where $b_{j,0}(x) \triangleq h_j(x)$. 

Next, for $i\in\{0,\ldots,d_j-2\}$ define
\begin{equation}\label{eq:C_j_i_def}
    \SC_{j,i} \triangleq \{x\in \BBR^n: b_{j,i}(x) \ge 0\}.
\end{equation}



For notational convenience, we define $b_j(x)  \triangleq b_{j, d_j-1}(x)$.
Then, consider $h:\BBR^n \to \BBR$ defined by
\begin{equation}\label{eq:h_def}
    h(x) \triangleq \mbox{softmin}_\rho \left ( b_1(x), b_2(x), \ldots,b_\ell(x) \right ), 
\end{equation}
and define 
\begin{equation}\label{eq:defS_softmin}
\SSS \triangleq \{ x \in \BBR^n \colon h(x) \ge 0 \}.
\end{equation}
The following result is the immediate consequence of Fact~\ref{fact:softmin_bound}.
\begin{fact}
$\SSS \subset \{x\in\BBR^n : b_1(x)\ge0,\ldots, b_\ell(x)\ge 0\}$.
\end{fact}

Next, define the sets
\begin{gather}  
    \SC_j \triangleq \bigcap_{i=0}^{d_j - 2} \SC_{j,i}, \qquad
    \SC \triangleq \bigcap_{j=1}^{\ell} \SC_j \subseteq \SSS_\rms \label{eq:C_def},
\end{gather}
and it follows that $\SSS \cap \SC \subset \SSS_\rms$. Furthermore, we note that if for all $x\in \mbox{bd } \SSS$, $L_gh(x) \neq 0$, then $\SSS \cap \SC$ is control forward invariant.

Next, we design a control that for all $t \ge 0$, seeks to minimize $J(x(t), u(t))$ subject to the constraint that $x(t) \in \SSS \cap \SC \cap \SSS_\rms$. Let $\gamma > 0$, and let $\alpha:\BBR \to \BBR$ be a nondecreasing function such that $\alpha(0) = 0$. 
For all $x \in \BBR^n$, the control is given by
\begin{subequations}\label{eq:qp_softmin}
\begin{align}
& \Big ( u(x), \mu(x) \Big ) \triangleq \underset{{\tilde u\in \BBR^m,\tilde \mu \in \BBR} }{\mbox{argmin}}  \, 
J(x, \tilde u) + \gamma \tilde \mu^2 \label{eq:qp_softmin.a}\\
& \text{subject to}\nn\\
& L_f h(x) + L_g h(x) \tilde u + \alpha (h(x)) + \tilde \mu h(x) \ge 0. \label{eq:qp_softmin.b}
\end{align}
\end{subequations}

The following result shows that the quadratic program~\eqref{eq:qp_softmin} is feasible if $L_gh(x) \neq 0$ for all $x \in \mbox{bd } \SSS$.

\begin{lemma}\label{lemma:qp_feas}\rm
Assume that for all $x\in \mbox{bd }\SSS$, $L_g h(x) \neq 0$. Then, for all $x\in\SSS$, there exists $\tilde u \in \BBR^m$ and $\tilde \mu \ge 0$ such that~\eqref{eq:qp_softmin.b} is satisfied.
\end{lemma}

The next theorem is the main result on the control~\Cref{eq:quad_cost,eq:b_def,eq:h_def,eq:qp_softmin} that uses the composite soft-minimum CBF $h$.

\begin{theorem}\label{thm:smocbf}\rm
Consider~\eqref{eq:dynamics}, where~\ref{cond:hocbf.a} and~\ref{cond:hocbf.b} are satisfied, and consider $u$ given by~\Cref{eq:quad_cost,eq:b_def,eq:h_def,eq:qp_softmin}.
Assume that for all $x \in \mbox{bd }\SSS, L_gh(x) \ne 0$. Let $x_0\in \SSS \cap \SC$.
Then, for all $t \ge 0 $, $x(t) \in \SSS\cap\SC \subset \SSS_\rms$.
\end{theorem}

Next, we present an example to demonstrate the control~\Cref{eq:quad_cost,eq:b_def,eq:h_def,eq:qp_softmin} that uses the composite soft-minimum CBF.

\begin{example}\label{example:nonholonomic_no_input_const}\rm
Consider the nonholonomic ground robot modeled by~\eqref{eq:dynamics}, where
\begin{equation*}
    f(x) = \begin{bmatrix}
    v \cos{\theta} \\
    v \sin{\theta} \\
    0 \\
    0
    \end{bmatrix}, 
    \,
    g(x) = \begin{bmatrix}
    0 & 0\\
    0 & 0\\
    1 & 0 \\
    0 & 1
    \end{bmatrix}, 
    \,
    x = \begin{bmatrix}
    q_\rmx\\
    q_\rmy\\
    v\\
    \theta
    \end{bmatrix}, 
    \,
    u = \begin{bmatrix}
    u_1\\
    u_2
    \end{bmatrix}, 
\end{equation*}
and $[ \, q_\rmx \quad q_\rmy \, ]^\rmT$ is the robot's position in an orthogonal coordinate system, $v$ is the speed, and $\theta$ is the direction of the velocity vector (i.e., the angle from $[ \, 1 \quad 0 \, ]^\rmT$ to $[ \, \dot q_\rmx \quad \dot q_\rmy \, ]^\rmT$).

Consider the map shown in~\Cref{fig:unicycle_hocbf_map}, which has 6 obstacles and a wall. 
For $j\in \{1,\ldots,6\}$, the area outside the $j$th obstacle is modeled as the $0$-superlevel set of 
\begin{equation*}\label{eq:h_obs}
h_j(x) = \left \| \matls a_{\rmx, j} (q_\rmx - b_{\rmx,j})  \\ a_{\rmy, j} (q_\rmy - b_{\rmy, j}) \matrs \right \|_p - c_j,    
\end{equation*}
where $b_{\rmx,j}, b_{\rmy,j}, a_{\rmx,j}, a_{\rmy,j}, c_j, p> 0$ specify the location and dimensions of the $j$th obstacle.
Similarly, the area inside the wall is modeled as the $0$-superlevel set of
$h_7(x) = c_7 - \left \| \matls a_{\rmx, 7} q_\rmx  \\ a_{\rmy, 7} q_\rmy \matrs \right \|_p$,
where $a_{\rmx, 7}, a_{\rmy, 7}, c_7 , p> 0$ specify the dimension of the space inside the wall. The bounds on speed $v$ are modeled as the $0$-superlevel sets of $h_8(x) = 9 - v$, $h_9(x) = v + 1$.
The safe set $\SSS_\rms$ is given by \eqref{eq:S_s} where $\ell = 9$, and its projection onto the $q_\rmx$--$q_\rmy$ plane is shown in~\Cref{fig:unicycle_hocbf_map}. Note that for all $x\in\SSS_\rms$, the speed satisfies $v\in[-1,9]$.

Let $q_{\rmd} = [\, q_{\rmd,\rmx} \quad q_{\rmd,\rmy}\,]^\rmT \in \BBR^2$ be the goal location, that is, the desired location for $[ \, q_\rmx \quad q_\rmy \, ]^\rmT$.

Next, the desired control is 
\begin{equation}\label{eq:u_d_ex}
    u_\rmd(x)\triangleq \begin{bmatrix}
    u_{\rmd_1}(x)\\u_{\rmd_2}(x)
\end{bmatrix},
\end{equation} 
where $u_{\rmd_1},u_{\rmd_2},r,\psi:\BBR^4 \to \BBR$ are defined by
\begin{align}
    u_{\rmd_1}(x) &\triangleq - (k_1 + k_3) v + (1+k_1 k_3) r(x) \cos \psi(x)\nn \\
    &\qquad + k_1 \left(k_2 r(x) + v\right) \sin^2 \psi(x), \label{eq:u_d_1_ex}\\
    u_{\rmd_2}(x) &\triangleq \left(k_2 + \frac{v}{r(x)}\right) \sin \psi(x), \label{eq:u_d_2_ex}\\
    r(x) &\triangleq \left \| \matls q_\rmx - q_{\rmd,\rmx}  \\ q_\rmy - q_{\rmd,\rmy} \matrs \right \|_2,\\
    \psi(x) &\triangleq \mbox{atan2}(q_\rmy-q_{\rmd,\rmy},q_\rmx-q_{\rmd,\rmx})-\theta + \pi,
\end{align} 
and $k_1 = 0.2, k_2= 1$, and $k_3 = 2$. Note that the desired control is designed using a process similar to \cite[pp. 30--31]{de2002control}, and it drives $[ \, q_\rmx \quad q_\rmy \, ]^\rmT$ to $q_\rmd$ but does not account for safety.

Next, we consider the cost function~\eqref{eq:quad_cost}, where $Q(x)= I_2$ and $c(x) = -u_\rmd(x)$. Thus, for each $x\in\BBR^n$, the minimizer of~\eqref{eq:quad_cost} is equal to the minimizer of the minimum-intervention cost $\|u - u_\rmd(x)\|_2^2$.

We note that~\ref{cond:hocbf.a} and~\ref{cond:hocbf.b} are satisfied by $d_1 = d_2 = \ldots = d_7 = 2$ and $d_8 = d_9 = 1$.

We implement the control~\Cref{eq:quad_cost,eq:b_def,eq:h_def,eq:qp_softmin} with $\rho = 10$, $\gamma = 10^{24}$, $\alpha_{1,0}(h) = \ldots= \alpha_{7,0}(h) = 7h$, and $\alpha(h(x)) = 0.5h(x)$. 
The control is updated at $1~\rm{kHz}$.

Figure~\ref{fig:unicycle_hocbf_map} shows the closed-loop trajectories for $x_0=[\,-1\quad -8.5\quad 0\quad \frac{\pi}{2}\,]^\rmT$ with 4 different goal locations $q_{\rmd}=[\,3\quad 4.5\,]^\rmT$, $q_{\rmd}=[\,-7\quad 0\,]^\rmT$, $q_{\rmd}=[\,7\quad 1.5\,]^\rmT$, and $q_{\rmd}=[\,-1\quad 7\,]^\rmT$. 
In all cases, the robot position converges to the goal location while satisfying safety constraints.

Figures~\ref{fig:unicycle_hocbf_states} and~\ref{fig:unicycle_hocbf_h} show the trajectories of the relevant signals for the case where $q_{\rmd}=[\,3 \quad 4.5\,]^\rmT$.
Figure~\ref{fig:unicycle_hocbf_h}
shows that $h$, $\min b_{j,i}$, and $\min h_{j}$ are positive for all time, which implies trajectory remains in $\SSS \cap \SC \subset \SSS_\rms$.
\exampletriangle
\begin{figure}[ht!]
\center{\includegraphics[width=0.38\textwidth,clip=true,trim= 0.2in 0.28in 1.2in 1.0in] {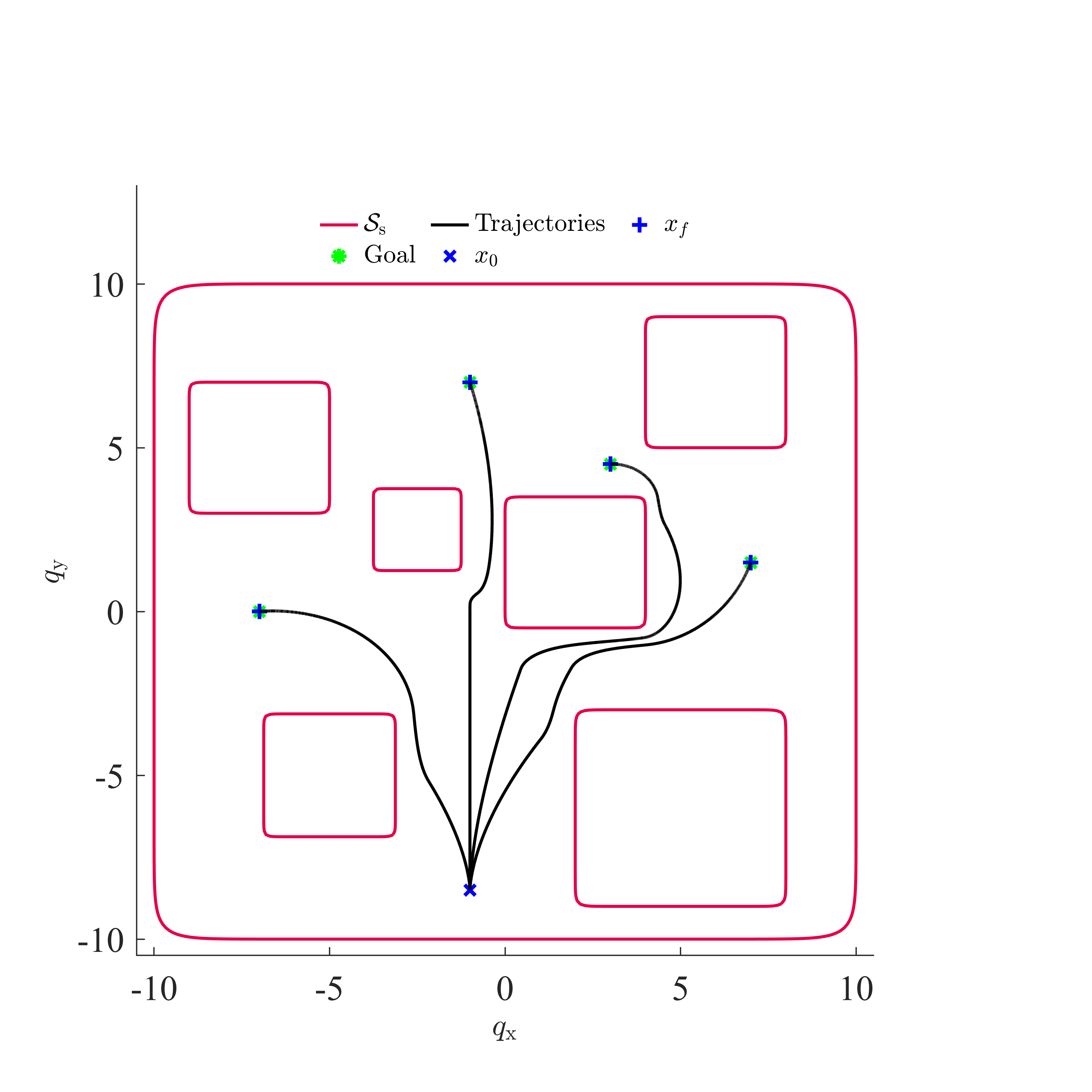}}
\caption{$\SSS_\rms$, and closed-loop trajectories for 4 goal locations.}\label{fig:unicycle_hocbf_map}
\end{figure}

\begin{figure}[h!]
\center{\includegraphics[width=0.47\textwidth,clip=true,trim= 0.3in 0.36in 1.1in 0.72in] {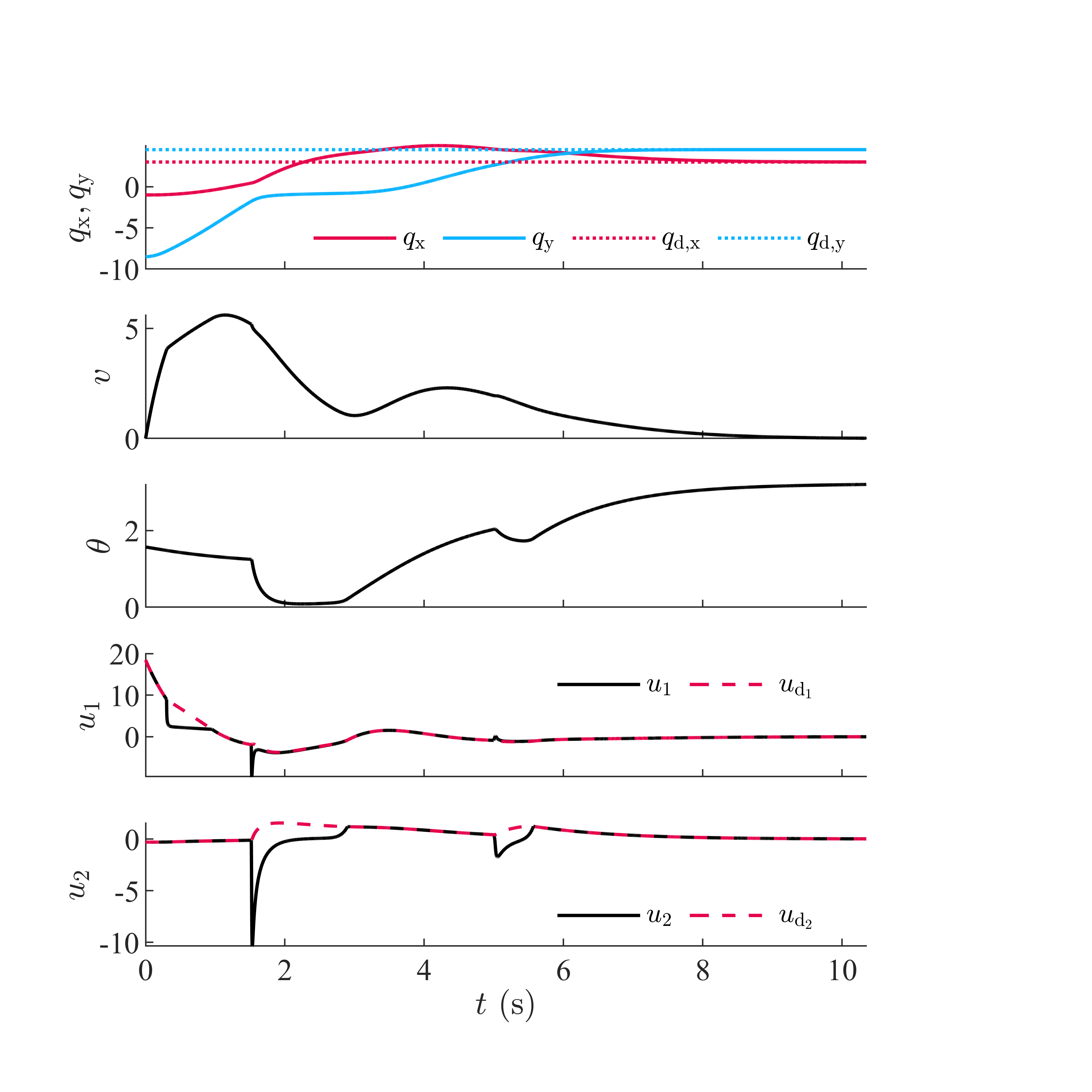}}
\caption{$q_\rmx$, $q_\rmy$, $v$, $\theta$, $u$ and $u_\rmd$ for $q_{\rmd}=[\,3 \quad 4.5\,]^\rmT$.}\label{fig:unicycle_hocbf_states}
\end{figure}

\begin{figure}[h!]
\center{\includegraphics[width=0.47\textwidth,clip=true,trim= 0.1in 0.25in 1.1in 0.6in] {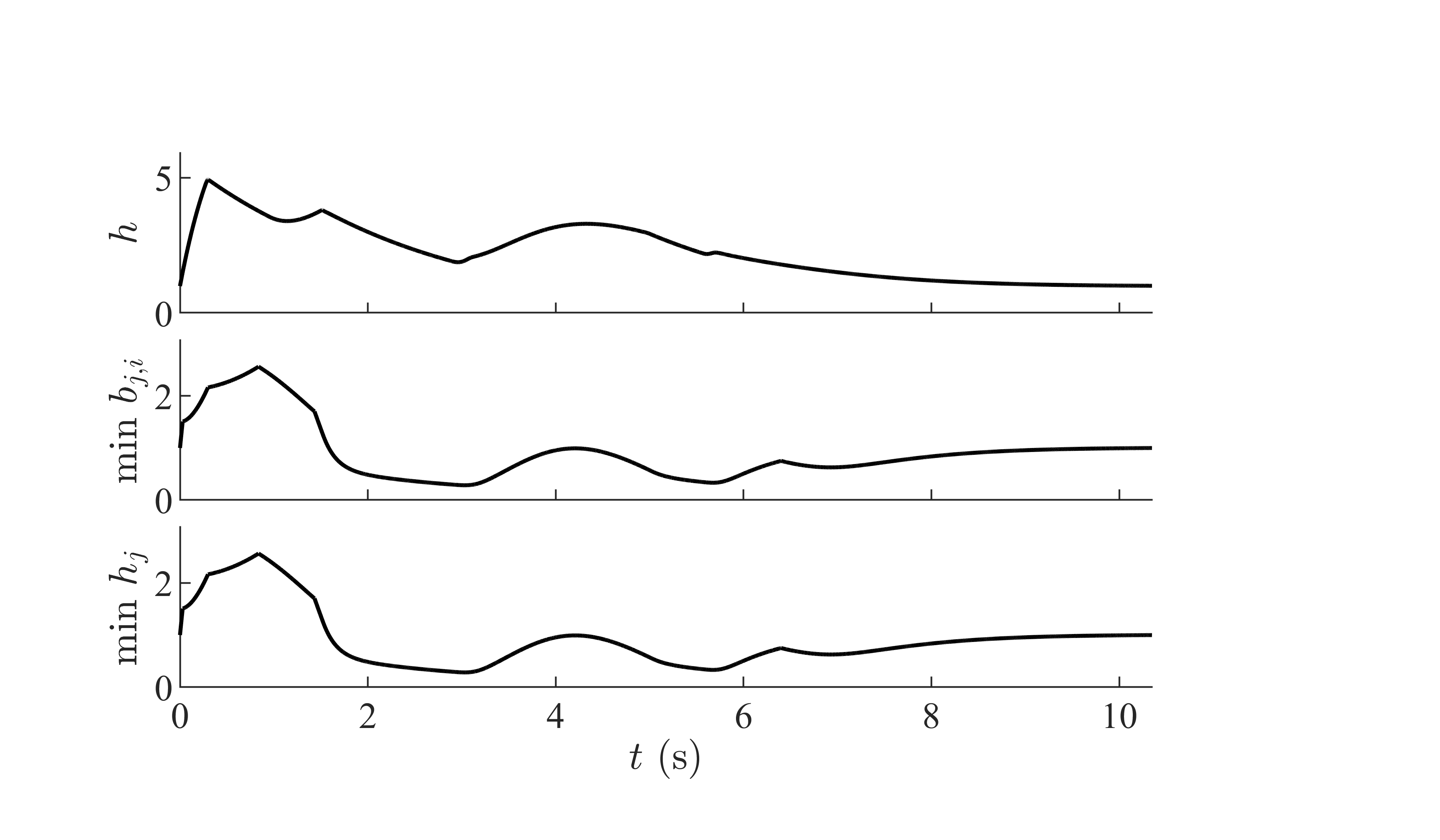}}
\caption{$h$, $\min b_{j,i}$, and $\min h_j$ for $q_{\rmd}=[\,3 \quad 4.5\,]^\rmT$.} \label{fig:unicycle_hocbf_h}
\end{figure}

\end{example}

\section{Guaranteed Safety with Input Constraints}
This section uses the composite soft-minimum CBF from the previous section to address safety with input constraints.

\subsection{Problem Formulation}
Consider
\begin{equation}\label{eq:dynamics_hat}
\dot {\hat x}(t) = \hat f(\hat x(t)) + \hat g(\hat x(t))\hat u(t),
\end{equation}
where $\hat x(t) \in \BBR^{\hat n}$ is the state, $\hat x(0) = \hat x_0 \in \BBR^{\hat n}$ is the initial condition, and $\hat u(t) \in \BBR^m$ is the control.

Let $\phi_1,\ldots,\phi_{\ell_{\hat u}}: \BBR^m \to \BBR$ be continuously differentiable, and define
\begin{align}\label{eq:U_def}
    \SU \triangleq \{\hat u\in \BBR^m: \phi_1(\hat u) \ge 0, \ldots,  \phi_{\ell_{\hat u}}(\hat u) \ge 0\} \subseteq \BBR^m,
\end{align}
which we assume is bounded and not empty.
Unless otherwise stated, all statements in this section that involve the subscript $\kappa$ are for all $\kappa \in \{1, 2, \ldots, \ell_{\hat u}\}$.
We assume that for all $\hat u \in \SU$, $\phi^\prime_\kappa(\hat u) \neq 0$.

Next, let $\hat h_{1},\ldots,h_{\ell_{\hat x}}:\BBR^{\hat n} \to \BBR$ be continuously differentiable, and define the safe set
\begin{equation}\label{eq:hat_S}
\hat \SSS_\rms \triangleq \{\hat x \in \BBR^{\hat n} \colon \hat h_{1}(\hat x)\geq 0, \hat h_2(\hat x)\ge 0 \ldots, \hat h_{\ell_{\hat x}}(\hat x) \geq 0 \}.
\end{equation}
Unless otherwise stated, all statements in this section that involve the subscript $\nu$ are for all $\nu \in \{ 1,2,\ldots,\ell_{\hat x}\}$.
We assume that there exists a positive integer $\hat d_\nu$ such that:
\begin{enumA}[resume]
    \item For all $\hat x \in \hat \SSS_\rms$ and all $i \in \{0, 1, \ldots, \hat d_{\nu}-2\}$, $L_{\hat g} L_{\hat f}^i \hat h_{\nu}(\hat x) = 0$.\label{cond:hocbf_x_hat.a}
    \item For all $\hat x\in \hat \SSS_\rms, L_{\hat g} L_{\hat f}^{\hat d_{\nu}-1}\hat h_{\nu}(\hat x)\neq 0$.\label{cond:hocbf_x_hat.b}
\end{enumA}

Next, consider the cost function $\hat J \colon \BBR^{\hat n} \times \BBR^m\to \BBR$ defined by
\begin{equation}\label{eq:quad_cost_hat}
    \hat J(\hat x, \hat u) \triangleq \frac{1}{2}\hat u^\rmT \hat Q(\hat x) \hat u + \hat c(\hat x)^\rmT \hat u,
\end{equation}
where $\hat Q:\BBR^{\hat n} \to \BBP^{m}$ and $c:\BBR^{\hat n}\to \BBR^m$.
The objective is to design a feedback control such that for all $t \ge 0$, $\hat J(\hat x(t), \hat u(t))$ is minimized subject to the safety constraints $x(t) \in \hat \SSS_\rms$ and the input constraint $\hat u(t) \in \SU$.

\subsection{Control for Safety with Input Constraints}
To address this problem, we introduce control dynamics. Specifically, consider a control $\hat u$ that satisfies
\begin{align}
    \dot x_\rmc(t) &= f_\rmc(x_\rmc(t)) + g_\rmc(x_\rmc(t))u(t),\label{eq:dynamics_control.a}\\
    \hat u(t) &= h_\rmc(x_\rmc(t)),\label{eq:dynamics_control.b}
\end{align}
where $x_\rmc(t) \in \BBR^{n_\rmc}$, $x_\rmc(0) = x_{\rmc0} \in \BBR^{n_\rmc}$ is the initial condition, and $u(t) \in \BBR^m$ is the input, which is computed from a quadratic program presented below.

Define the set
\begin{equation*}\label{eq:S_c_def}
\SSS_{\rmc} \triangleq \{x_\rmc \in\BBR^{n_\rmc}:h_\rmc(x_\rmc) \in \SU\}.
\end{equation*}
%
The functions $f_\rmc$, $g_\rmc$ and $h_\rmc$ are selected such that there exist positive integers $d_\rmc$ and $\zeta$ such that the following conditions hold:
\begin{enumC}
    \item For all $x_\rmc \in \SSS_{\rmc}$ and all $i \in \{0, 1, \ldots, d_\rmc-2\}$, $L_{g_\rmc} L_{f_\rmc}^i h_\rmc(x_\rmc) = 0$.\label{cond:hocbf_x_c.a}
    \item For all $x_\rmc \in \SSS_{\rmc}$, $L_{g_\rmc} L_{f_\rmc}^{d_\rmc-1}h_\rmc(x_\rmc)$ is nonsingular.\label{cond:hocbf_x_c.b}
    \item For all $\hat x \in \hat \SSS_\rms$ and all $x_\rmc \in \SSS_{\rmc}$, $L_{\hat g}L_{\hat f}^{\hat d_\nu - 1}{\hat h_\nu}(\hat x)L_{g_\rmc}L_{f_\rmc}^{d_\rmc-1}h_\rmc(x_\rmc) \neq 0$.\label{cond:hocbf_x_c.c}
    \item For all $x_\rmc \in \SSS_{\rmc}$ and all $i \in \{0, 1, \ldots, \zeta-2\}$, $L_{g_\rmc} L_{f_\rmc}^i \phi_\kappa(h_\rmc(x_\rmc)) = 0$.\label{cond:hocbf_x_c.d}
    \item For all $x_\rmc \in \SSS_{\rmc}$, $L_{g_\rmc} L_{f_\rmc}^{\zeta-1}\phi_\kappa(h_\rmc(x_\rmc)) \neq 0$. \label{cond:hocbf_x_c.e}
\end{enumC}

Note that there exist constructions of $f_\rmc$, $g_\rmc$, and $h_\rmc$ for which~\ref{cond:hocbf_x_c.a}--\ref{cond:hocbf_x_c.e} are satisfied. The following example provides one such construction.

\begin{example}
\rm{
Let 
\begin{gather}\label{eq:lti_sys}
f_\rmc(x_\rmc) = A_\rmc x_\rmc, \quad g_\rmc(x_\rmc) = B_\rmc, \quad h_\rmc(x_\rmc) = C_\rmc x_\rmc, 
\end{gather}
where $A_\rmc \in \BBR^{n_\rmc \times n_\rmc}$, $B_\rmc \in \BBR^{n_\rmc \times m}$, $C_\rmc \in \BBR^{m\times n_\rmc}$, and $C_\rmc B_\rmc$ is nonsingular. 
Thus, $L_{g_\rmc}h_\rmc(x_\rmc) = C_\rmc B_\rmc$ is nonsingular, which implies that \ref{cond:hocbf_x_c.a} and \ref{cond:hocbf_x_c.b} are satisfied with $d_\rmc = 1$. 
Next, since $L_{g_\rmc}h_\rmc(x_\rmc)$ is nonsingular, it follows from \ref{cond:hocbf.b} that \ref{cond:hocbf_x_c.c} is satisfied. 
Finally, note that $L_{g_\rmc}\phi_\kappa(h_\rmc(x_\rmc)) = \phi_\kappa^\prime(h_\rmc(x_\rmc) L_{g_\rmc}h_\rmc(x_\rmc) = \phi_\kappa^\prime(h_\rmc(x_\rmc)) C_\rmc B_\rmc$.
Since $C_\rmc B_\rmc$ is nonsingular and for all $x_\rmc \in \SSS_\rmc$, 
$\phi_\kappa^\prime(h_\rmc(x_\rmc))\neq 0$, it follows that \ref{cond:hocbf_x_c.d} and \ref{cond:hocbf_x_c.e} are satisfied with $\zeta= 1$. 
    }
\exampletriangle
\end{example}

Next, the cascade of~\cref{eq:dynamics_hat,eq:dynamics_control.a,eq:dynamics_control.b} is given by \eqref{eq:dynamics}, where
\begin{align}\label{eq:dynamics_aug}
        f(x) &\triangleq \begin{bmatrix}
            \hat f(\hat x) + \hat g(\hat x)h_\rmc(x_\rmc)\\
            f_\rmc(x_\rmc)
        \end{bmatrix},
        g(x) \triangleq \begin{bmatrix}
            0\\
            g_\rmc(x_\rmc)
        \end{bmatrix}, x \triangleq  \begin{bmatrix}
            \hat x \\ x_\rmc
        \end{bmatrix},
\end{align}
and $n\triangleq \hat n + n_\rmc$.
Let $\ell \triangleq \ell_{\hat x} + \ell_{\hat u}$, and for $j\in\{1,2, \ldots, \ell\}$, $h_j$ is defined as
\begin{align}\label{eq:h_j_def}
    h_j(x) \triangleq \begin{cases}
        \hat h_j(\hat x), & \mbox{if } j\in\{1,2,\ldots, \ell_{\hat x}\},\\
        \phi_{j - \ell_{\hat x}}(h_\rmc(x_\rmc)), & \mbox{if } j\in\{\ell_{\hat x}+1,\ell_{\hat x}+2,\ldots, \ell\}.
    \end{cases}
\end{align}
Consider the set $\SSS_\rms$ defined in~\eqref{eq:S_s}, and note that $\SSS_\rms = \hat \SSS_\rms \times \SSS_{c}$. The following result demonstrates that the cascade~\eqref{eq:dynamics_aug} satisfies~\ref{cond:hocbf.a} and~\ref{cond:hocbf.b}.

\begin{proposition}\label{prop:cascade}\rm
Consider~\eqref{eq:dynamics_hat}, where~\ref{cond:hocbf_x_hat.a} and~\ref{cond:hocbf_x_hat.b} are satisfied, and consider \cref{eq:dynamics_control.a,eq:dynamics_control.b}, where~\ref{cond:hocbf_x_c.a}--\ref{cond:hocbf_x_c.e} are satisfied.
Then, $f$, $g$, and $h_j$ given by \cref{eq:dynamics_aug,eq:h_j_def} satisfy~\ref{cond:hocbf.a} and~\ref{cond:hocbf.b} with 
\begin{align}\label{eq:d_j_def}
    d_j \triangleq \begin{cases}
        \hat d_j + d_\rmc, & \mbox{if } j\in\{1,2,\ldots, \ell_{\hat x}\},\\
        \zeta, & \mbox{if } j\in\{\ell_{\hat x}+1,\ell_{\hat x}+2,\ldots, \ell\}.
    \end{cases}
\end{align}    

\end{proposition}


Since~\Cref{prop:cascade} implies that the cascade~\eqref{eq:dynamics_aug} with the CBFs~\eqref{eq:h_j_def} satisfy~\ref{cond:hocbf.a} and~\ref{cond:hocbf.b}, it follows that the composite soft-minimum CBF construction from the previous section can be applied to combine the CBFs~\eqref{eq:h_j_def} that describe a set $\SSS_\rms = \hat \SSS_\rms \times \SSS_\rmc$ that combines the safe set $\hat \SSS_\rms$ and the set of controller states $\SSS_\rmc$ that yield control signals in the admissible set $\SU$. In other words, we can apply the composite soft-minimum CBF construction to enforce safety and input constraints. To do this, for $i\in\{0, \ldots, d_j -2\}$, let $b_{j,i+1}:\BBR^n \to \BBR$ be given by~\eqref{eq:b_def}. Furthermore, let $h:\BBR^n \to \BBR$ be the composite soft-minimum CBF given by~\cref{eq:h_def}, and let $\SSS$ be given by~\cref{eq:defS_softmin}.

Next, note that we cannot directly apply the quadratic program~\eqref{eq:qp_softmin} because the cost $J$ given by~\eqref{eq:quad_cost} is a function of $u$, whereas the cost $\hat J$ given by~\eqref{eq:quad_cost_hat} is a function of $\hat u$. Thus, we introduce an expression for the cost $J$ that is a surrogate for the cost $\hat J$. In other words, we develop an expression for $J$ such that minimizing $J$ tends to minimize $\hat J$.


Consider $\hat u_\rmd:\BBR^{n} \to \BBR^m$ defined by
\begin{align}\label{eq:u_d_hat_def}
\hat u_\rmd(x) \triangleq - \hat Q(\hat x)^{-1} \hat c(\hat x),
\end{align}
which is the minimizer of~\eqref{eq:quad_cost_hat}.
Next, let $\gamma_{d_\rmc} = 1$, and consider $u_\rmd: \BBR^n \to \BBR^m$ defined by
\begin{align}\label{eq:u_d_def}
    u_\rmd(x) \triangleq &\left(L_{g_\rmc} L_{f_\rmc}^{d_\rmc - 1} h_\rmc(x_\rmc)\right)^{-1} \Bigg(\sum_{i = 0}^{d_\rmc} \gamma_i \Big(L_f^{i}\hat u_\rmd(x)\nn\\
    &\qquad -L_{f_\rmc}^i h_\rmc(x_\rmc)\Big)\Bigg),
\end{align}
where $\gamma_0, \gamma_1, \ldots, \gamma_{d_\rmc -1}>0$ are selected such that 
\begin{align}\label{eq:char_poly}
s^{d_\rmc}+ \gamma_{d_\rmc -1} s^{d_\rmc -1} + \gamma_{d_\rmc -2} s^{d_\rmc - 2}+ \cdots + \gamma_1 s + \gamma_0
\end{align}
has all its roots in the open left-hand complex plane.



The following result considers the closed-loop~\cref{eq:dynamics_hat,eq:dynamics_control.a,eq:dynamics_control.b} system under $u_\rmd$.
This result demonstrates that the trajectory of closed-loop system under $u_\rmd$ converges exponentially to the trajectory of~\eqref{eq:dynamics_hat} under the ideal control $\hat u_\rmd$, which minimizes $\hat J$.

\begin{proposition}\label{proposition:error}
\rm{
Consider~\cref{eq:dynamics_hat,eq:dynamics_control.a,eq:dynamics_control.b}, where $u=u_\rmd$. 
For all $t\ge 0$, define $e(t)\triangleq \hat u(t) - \hat u_\rmd(x(t))$. Then, the following statements hold:
\begin{enumalph}
%

%
\item The signal $e$ satisfies
$e^{(d_\rmc)} + \gamma_{d_\rmc-1} e^{(d_\rmc-1)} + \cdots + \gamma_1 \dot e + \gamma_0 e = 0$, and for all $x(0) \in \BBR^n$, $\lim_{t \to \infty} e(t) =0$ exponentially.
\label{proposition:error.b}
\item Assume that $\hat Q(\hat x)$ is bounded. 
Then, for all $x(0) \in \BBR^n$, 
    \begin{equation*}
    \lim_{t \to \infty} \left[\hat J(\hat x(t), \hat u(t)) - \hat J(\hat x(t), \hat u_\rmd(x(t)))\right] = 0.
    \end{equation*}
\label{proposition:error.c}
%
%
\item Let $\hat x_0\in \BBR^{\hat n}$.
Assume that $x_{\rmc 0}\in \BBR^{n_\rmc}$ is such that for $i \in \{0,1,\ldots, d_\rmc -1\}$,  $L_f^i h_\rmc(x_{\rmc 0}) = L_f^i \hat u_\rmd(x_0)$, where $x_0 \triangleq [\hat x_0^\rmT \quad  x_{\rmc 0}^\rmT]^\rmT$. Then, for all $t\ge 0$, $e(t) = 0$.
\label{proposition:error.d}

\end{enumalph}
}
\end{proposition}

\Cref{proposition:error} implies that $u=u_\rmd$ yields an output $\hat u$ from the control dynamics~\cref{eq:dynamics_control.a,eq:dynamics_control.b} that converges exponentially to the minimizer $\hat u_\rmd$ of $\hat J$. Thus, we consider a cost $J$ that is minimized by $u_\rmd$. Specifically, consider $J$ given by~\eqref{eq:quad_cost}, where
\begin{equation}\label{eq:cost_Q_c_def}
Q(x) = I_n,\quad c(x) =-u_\rmd(x).
\end{equation}
Finally, we consider the input signal $u$ to the controller dynamics~\cref{eq:dynamics_control.a,eq:dynamics_control.b} given by~\eqref{eq:qp_softmin}, where $\gamma >0$ and $\alpha:\BBR\to\BBR$ is a nondecreasing function with $\alpha(0) = 0$. 





The following corollary is the main result on the control~\Cref{eq:quad_cost,eq:b_def,eq:h_def,eq:qp_softmin,eq:dynamics_control.a,eq:dynamics_control.b,eq:dynamics_aug,eq:h_j_def,eq:u_d_hat_def,eq:u_d_def,eq:cost_Q_c_def} that uses the composite soft-minimum CBF.

\begin{corollary}\label{cor:input_const}
\rm{
Consider~\eqref{eq:dynamics_hat}, where~\ref{cond:hocbf_x_hat.a} and~\ref{cond:hocbf_x_hat.b} are satisfied.
Furthermore, consider the control $\hat u$ is given by~\cref{eq:dynamics_control.a,eq:dynamics_control.b}, where~\ref{cond:hocbf_x_c.a}--\ref{cond:hocbf_x_c.e} are satisfied, and $u$ is given by \cref{eq:quad_cost,eq:b_def,eq:h_def,eq:qp_softmin,eq:dynamics_aug,eq:h_j_def,eq:u_d_hat_def,eq:u_d_def,eq:cost_Q_c_def}.
Assume that for all $x \in \mbox{bd }\SSS, L_gh(x) \ne 0$. 
Let $x_0\in \SSS \cap \SC$.
Then, for all $t \ge 0 $, $x(t) \in \SSS\cap\SC$, $\hat x(t) \in \hat \SSS_\rms$, and $\hat u(t) \in \SU$.
}
\end{corollary}

Next, we present an example that demonstrates the composite soft-minimum CBF approach for safety with input constraints. 

\begin{example}\rm
We revisit the nonholonomic ground robot from~\Cref{example:nonholonomic_no_input_const} and include not only safety constraints but also input constraints. The nonholonomic ground robot modeled by~\eqref{eq:dynamics_hat}, where
\begin{equation*}
    \hat f(\hat x) = \begin{bmatrix}
    v \cos{\theta} \\
    v \sin{\theta} \\
    0 \\
    0
    \end{bmatrix}, 
    \,
    \hat g(\hat x) = \begin{bmatrix}
    0 & 0\\
    0 & 0\\
    1 & 0 \\
    0 & 1
    \end{bmatrix}, 
    \,
    \hat x = \begin{bmatrix}
    q_\rmx\\
    q_\rmy\\
    v\\
    \theta
    \end{bmatrix}, 
    \,
    \hat u = \begin{bmatrix}
    \hat u_1\\
    \hat u_2
    \end{bmatrix}, 
\end{equation*}
and $q_\rmx, q_\rmy, v, \theta$ are the same as in~\Cref{example:nonholonomic_no_input_const}. 

Consider the map shown in~\Cref{fig:unicycle_hocbf_map_input_constraint}, 
which is the same map considered in~\Cref{example:nonholonomic_no_input_const}.
For $j\in \{1,\ldots,9\}$, consider $\hat h_j(\hat x)$ given by the expression for $h_j(x)$ in \Cref{example:nonholonomic_no_input_const}. The safe set $\hat \SSS_\rms$ is given by \eqref{eq:hat_S}, and its projection onto the $q_\rmx$--$q_\rmy$ plane is shown in~\Cref{fig:unicycle_hocbf_map_input_constraint}.
The bounds on input $\hat u$ are modeled as the $0$-superlevel sets of 
\begin{align*}
    \phi_1(\hat u) &= 4 - \hat u_1, \qquad
    \phi_2(\hat u) = \hat u_1 + 4,\\
    \phi_3(\hat u) &= 1 - \hat u_2,\qquad
    \phi_4(\hat u) = \hat u_2 + 1,
\end{align*}
and $\SU$ is given by~\eqref{eq:U_def}.

Let $q_{\rmd} = [\, q_{\rmd,\rmx} \quad q_{\rmd,\rmy}\,]^\rmT \in \BBR^2$ be the goal location, that is, the desired location for $[ \, q_\rmx \quad q_\rmy \, ]^\rmT$.
Next, we consider a desired control $\hat u_\rmd (\hat x)$ that is the same as $u_\rmd(x)$ considered in Example 1.
Specifically, $\hat u_\rmd$, $\hat u_{\rmd_1} $, and $\hat u_{\rmd_2}$ are given by~\cref{eq:u_d_ex,eq:u_d_1_ex,eq:u_d_2_ex}, where
$u_\rmd$, $u_{\rmd_1}$ and $u_{\rmd_2}$ are replaced by $\hat u_\rmd$, $\hat u_{\rmd_1}$ and $\hat u_{\rmd_2}$. Thus, we consider the cost function~\eqref{eq:quad_cost_hat}, where $\hat Q(x)= I_2$ and $\hat c(x) = -\hat u_\rmd(x)$.

We consider the controller dynamics~\cref{eq:dynamics_control.a,eq:dynamics_control.b,eq:lti_sys}, where
$A_\rmc = -I_2$, $B_\rmc = I_2$, $C_\rmc = I_2$, $u=[\,u_1\quad u_2\,]^\rmT$.

Let $f$ and $g$ be given by~\eqref{eq:dynamics_aug}. We note that~\ref{cond:hocbf.a} and~\ref{cond:hocbf.b} are satisfied by $d_1 = d_2 = \ldots = d_7 = 3$, $d_8 = d_9 = 2$, $d_{10} = d_{11} = d_{12} = d_{13} = 1$, $d_\rmc = 1$, $\zeta = 1$. 

We implement the control~\Cref{eq:quad_cost,eq:b_def,eq:h_def,eq:qp_softmin,eq:dynamics_control.a,eq:dynamics_control.b,eq:dynamics_aug,eq:h_j_def,eq:u_d_hat_def,eq:u_d_def,eq:cost_Q_c_def} with $\rho = 10$, $\gamma = 100$, $\alpha_{1,0}(h) = \ldots= \alpha_{6,0}(h) = 1$, $\alpha_{1,1}(h) = \ldots= \alpha_{6,1}(h) = 2.5h$, $\alpha_{7,0}(h) = 6$, $\alpha_{7,1}(h) = 1$, $\alpha_{8,0}(h) = \alpha_{9,0}(h) = 10h$, $\alpha = 0$, and $\gamma_1 = 1$. The control is updated at $1~\rm{kHz}$.

Figure~\ref{fig:unicycle_hocbf_map_input_constraint} shows the closed-loop trajectories for $x_0=[\,-1\quad -8.5\quad 0\quad \frac{\pi}{2}\quad0\quad0\,]^\rmT$
with 4 different  goal locations $q_{\rmd}=[\,3\quad 4.5\,]^\rmT$, $q_{\rmd}=[\,-7\quad 0\,]^\rmT$, $q_{\rmd}=[\,7\quad 1.5\,]^\rmT$, and $q_{\rmd}=[\,-1\quad 7\,]^\rmT$.
In all cases, the robot position converges to the goal location while satisfying safety and input constraints.

Figures~\ref{fig:unicycle_hocbf_states_input_constraint} and~\ref{fig:unicycle_hocbf_h_input_constraint} show the trajectories of the relevant signals for the case where $q_{\rmd}=[\,3 \quad 4.5\,]^\rmT$.
Figure~\ref{fig:unicycle_hocbf_h_input_constraint}
shows that $h$, $\min b_{j,i}$, and $\min h_{j}$ are positive for all time, which implies  that $\hat x$ remains in $\hat \SSS_\rms$ and $\hat u$ remains in $\SU$.
\exampletriangle

\begin{figure}[ht!]
\center{\includegraphics[width=0.38\textwidth,clip=true,trim= 0.2in 0.23in 1.2in 1.0in] {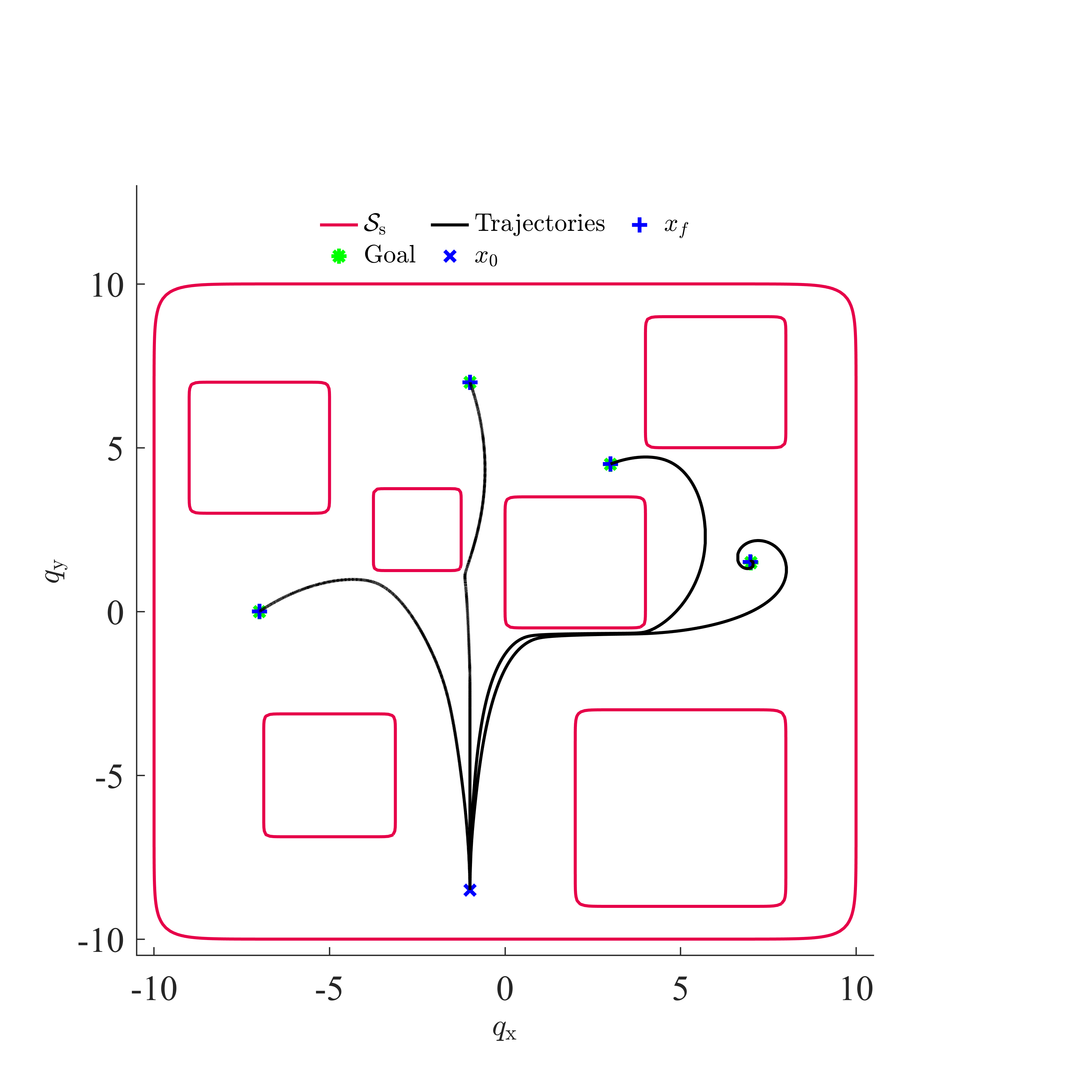}}
\caption{$\SSS_\rms$ and closed-loop trajectories for 4 goal locations.}\label{fig:unicycle_hocbf_map_input_constraint}
\end{figure}

\begin{figure}[h!]
\center{\includegraphics[width=0.47\textwidth,clip=true,trim= 0.3in 0.4in 1.1in 0.75in] {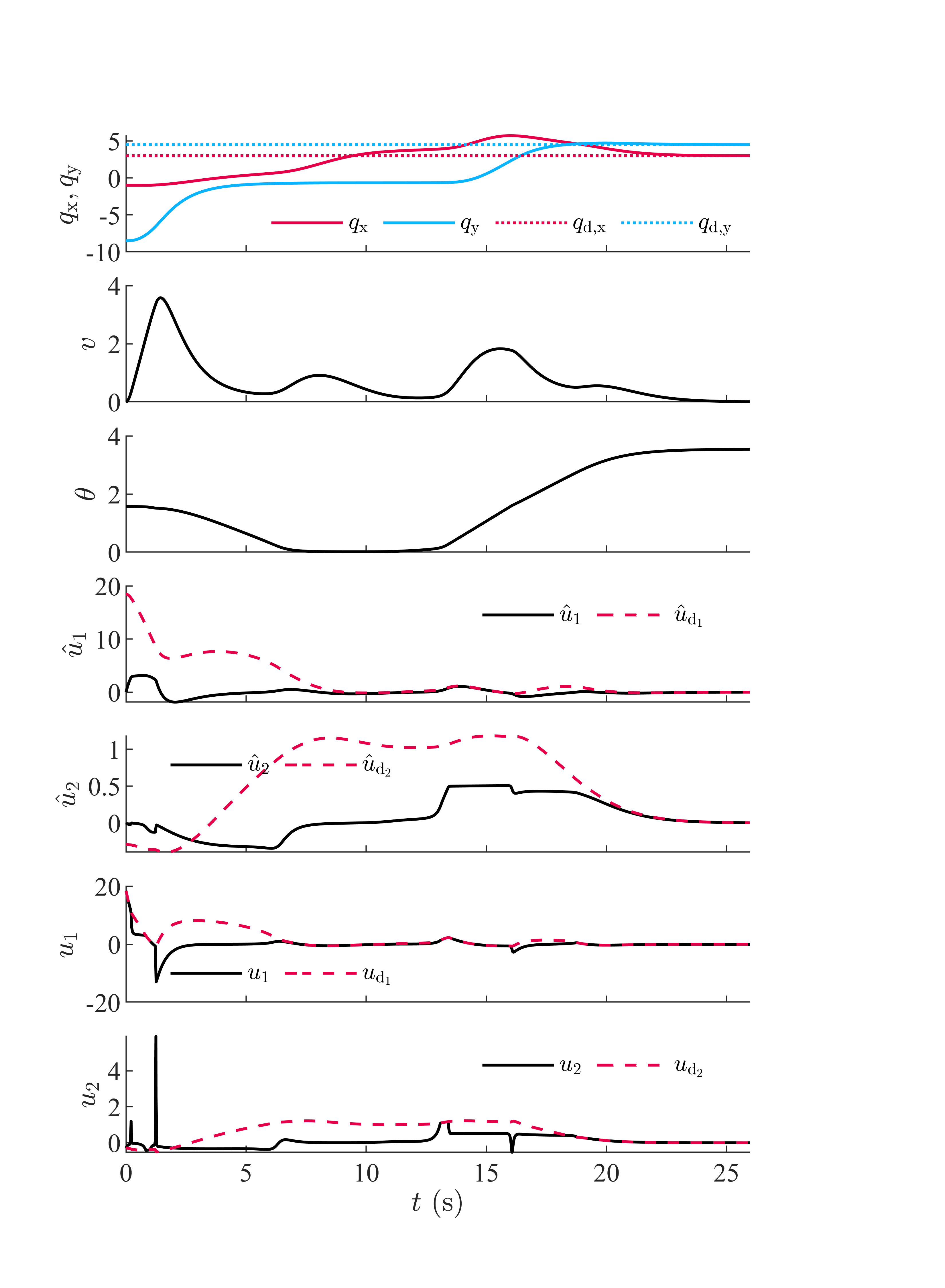}}
\caption{$q_\rmx$, $q_\rmy$, $v$, $\theta$, $\hat u$, $\hat u_\rmd$, $e$, $u$ and $u_\rmd$ for $q_{\rmd}=[\,3 \quad 4.5\,]^\rmT$.}\label{fig:unicycle_hocbf_states_input_constraint}
\end{figure}

\begin{figure}[h!]
\center{\includegraphics[width=0.47\textwidth,clip=true,trim= 0.15in 0.25in 1.1in 0.55in] {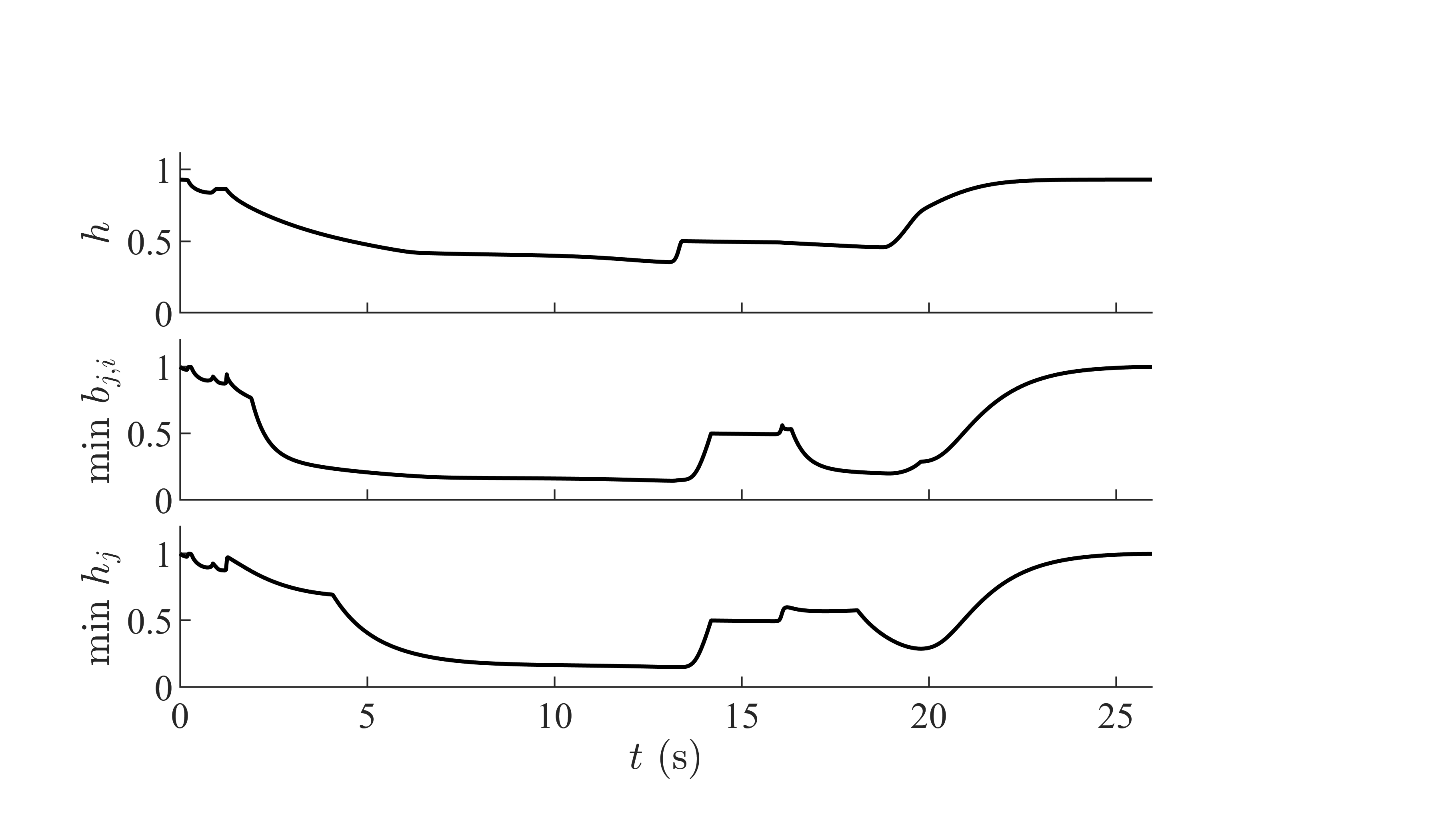}}
\caption{$h$, $\min b_{j,i}$, and $\min h_j$ for $q_{\rmd}=[\,3 \quad 4.5\,]^\rmT$.} \label{fig:unicycle_hocbf_h_input_constraint}
\end{figure}

\end{example}

\appendices
\gdef\thesection{\Alph{section}}

 \bibliographystyle{elsarticle-num} 
 \bibliography{softmin_hocbf}
 
\end{document}